\documentclass[reprint, twocolumn, superscriptaddress,prb]{revtex4-2}
\usepackage{bm, amsmath, amsfonts, amssymb, braket}
\usepackage{subfigure}
\usepackage{color}
\usepackage{graphicx}
\usepackage{slashed}

\usepackage{dcolumn} 

\usepackage{rotating} 

\usepackage{ulem}
\usepackage{xcolor}
\DeclareRobustCommand{\erase}{\bgroup\markoverwith{\textcolor{red}{\rule[.5ex]{2pt}{0.4pt}}}\ULon}

\begin{document}
\title{Constructing vortex functions and basis states of Chern insulators: ideal condition, inequality from index theorem, and coherent-like states on von Neumann lattice}
\author{Nobuyuki Okuma}
\email{okuma@hosi.phys.s.u-tokyo.ac.jp}
\affiliation{%
  Graduate School of Engineering, Kyushu Institute of Technology, Kitakyushu 804-8550, Japan
}%

\date{\today}
\begin{abstract}
In the field of fractional Chern insulators, a great deal of effort has been devoted to characterizing Chern bands that exhibit properties similar to the Landau levels.
Among them, the concept of the vortex function, which generalizes the complex coordinate used for the symmetric-gauge Landau-level basis, allows for a concise description.
In this paper, we develop a theory of constructing the vortex function and basis states of Chern insulators in the tight-binding formalism.
In the first half, we consider the optimization process of the vortex function, which minimizes an indicator that measures the difference from the ideal Chern insulators.
In particular, we focus on the sublattice position dependence of the vortex function or the quantum geometric tensor.
This degree of freedom serves as a discrete analog of the non-uniformity in the spatial metric and magnetic field in a continuous model.
In the second half, we construct two types of basis sets for a given vortex function: radially localized basis set and coherent-like basis set.
The former basis set is defined as the eigenstates of an analogy of the angular momentum operator.
Remarkably, one can always find exact zero mode(s) for this operator, which is explained by the celebrated Atiyah-Singer index theorem.
As a byproduct, we propose an inequality rooted in the band topology.
We also discuss the subtle differences between our formalism and the previous works about the momentum-space Landau level.
The latter basis set generalizes the concept of coherent states on von Neumann lattice. 
While this basis set is not orthogonal, it is useful to compare the LLL and the given Chern insulator directly in the Brillouin zone. 
These basis sets are expected to be useful for many-body calculations of fractional Chern insulators.

\end{abstract}
\maketitle
\section{Introduction}
The physics of topological phases has attracted much attention in recent years \cite{Kane-review,Zhang-review}.
Among them, the integer \cite{Klitzing-80} and fractional Quantum Hall Effects (QHEs) \cite{FQHE-exp-82,Laughlin-Wavefunction-83,yoshioka-textbook-02,halperin-fractional-textbook-20,fradkin2013field} are positioned as pioneering and fundamental examples. 
Despite involving interesting phenomena such as quantized conductivity and the appearance of robust edge states \cite{Harperin-82,Hatsugai-93-PRB,Hatsugai-93-PRL}, the integer QHE is essentially explained in terms of the one-particle picture. Inspired by the simple physics of the integer QHE, the notion of topological insulators has been established as a broader concept \cite{Kane-review,Zhang-review}.
In contrast, the fractional QHE requires the interplay between bulk topology and strong correlation \cite{FQHE-exp-82,Laughlin-Wavefunction-83,yoshioka-textbook-02,halperin-fractional-textbook-20}.
Mathematically, the fractional QHE is classified as an intrinsic topological order \cite{wen1990topological}, 
a concept that defies understanding through a one-particle picture.
Unlike topological insulators, which lack bulk degrees of freedom, systems with topological order harbor fractional excitations within the bulk \cite{Laughlin-anyon-83}, known as anyons \cite{leinaas1977theory,wilczek1982quantum}.

While the QHE requires a strong magnetic field, the topological insulators can be defined in the absence of external fields.
The direct counterpart of the integer QHE is called a Chern insulator~\cite{Haldane-88}, which is now classified as an example of a two-dimensional topological insulator.
Instead of relying on magnetic fields, Chern insulators use the nontrivial Chern number inherent in the occupied bands to achieve the same transport properties as the integer QHE.
Similarly, lattice implementations of the fractional QHE, called fractional Chern insulators (FCIs) \cite{Regnault-Bernevig-11, Bergholtz-Liu-13, Parameswaran-13, Liu-Bergholtz-review-22}, have been extensively studied for interacting Hamiltonians with a fractional filling whose non-interacting part is a Chern insulator.
Unlike the Chern insulator, however, the realization of the FCI is a nontrivial task.
Theoretical aspects of the fractional QHE are sufficiently established only for the Landau levels.
In particular, the flatness of the one-particle energy is important for relatively strong correlation effects, and the analytic properties of Landau levels are essential for the construction of many-body states such as the Laughlin wavefunction \cite{Laughlin-Wavefunction-83}.
The Chern insulator generally does not share other properties of the Landau levels beyond the topological characteristics.
Therefore, efforts in this field have been directed towards characterizing the Chern bands, which exhibit properties akin to Landau levels \cite{Regnault-Bernevig-11, Bergholtz-Liu-13, Parameswaran-13, Liu-Bergholtz-review-22,Parameswaran-Roy-Sondhi-12,Roy-geometry-14,Jackson-Moller-Roy-15,Claassen-Lee-Thomale-Qi-Devereaux-15,Lee-Claassen-Thomale-17,Mera-Ozawa-21-2,Varjas-Abouelkomsan-Kang-Bergholtz-22}.
Despite these theoretical difficulties, experimental papers claiming to have achieved FCI in bilayer systems have also emerged, attracting sudden attention \cite{li2021spontaneous,cai2023signatures,zeng2023thermodynamic,park2023observation,xu2023observation,lu2023fractional}.

Very recently, a concept called ``vortexability" has been introduced \cite{ledwith2023vortexability}, which includes and generalizes previous studies in this area.
In the lowest Landau level (LLL), the operation of multiplying by a complex coordinate $z:=x+iy$ remains closed within the LLL. Physically, this process is interpreted as an attachment of a vortex with the flux quantum.
Reference \cite{ledwith2023vortexability} generalized the concept of complex coordinate function in the LLL to the Chern insulator, naming it the vortex function $z(\bm{r})$.
When the operation of multiplying by $z(\bm{r})$ is closed in the Chern band, it is called a vortexable band.
The concept of vortexability is not only practical but also facilitates a concise explanation of the complicated theory of the FCI.

This paper conducts a study on the topics concerning the vortex function in the lattice models of Chern bands.
As Chern bands are not universally vortexable, we explore an optimal function that resembles the vortex function according to certain criteria.
By broadening the definition, we categorize these functions as vortex functions.
Our objective is to construct optimal vortex functions and practical basis states for Chern-insulator tight-binding models.

This paper is organized as follows.
In Sect.\ref{optimize}, we consider the problem of optimizing the vortex function for a given Chern band.
In particular, we focus on the sublattice-position-dependence of the vortex function and consider the optimization process with the virtual lattice reconstruction. 
We investigate several models of Chern insulators and evaluate their vortexability by using an indicator that is defined for a given vortex function. 
In Sect.\ref{radialbasis} and Sect.\ref{clsvnl}, we consider two types of basis states for a given Chern band. 
In Sect.\ref{radialbasis}, we consider a generalization of the Landau-level radially localized basis under the symmetric gauge.
We construct the radially localized basis states as the eigenstates of an analogy of the angular momentum operator defined within the Chern band.
Remarkably, one can always find the exact zero mode(s), which is explained by the celebrated Atiyah-Singer index theorem in momentum space. Related to these zero modes, we propose an inequality rooted in nontrivial topology.
We also discuss subtle differences between our basis set and the previous study of the momentum-space Landau level \cite{Claassen-Lee-Thomale-Qi-Devereaux-15,Lee-Claassen-Thomale-17}.
In Sect.\ref{clsvnl}, we consider a generalization of the coherent states on the Landau levels.
We construct non-orthogonal localized states on unit cells, which resemble the complete set of coherent states on the von Neumann lattice. This formalism enables us to compare the LLL and the given Chern insulator directly in the Brillouin zone. 
\\
\textcolor{red}{Note added: the red text indicates the parts that were corrected after publication in Physical Review B. }

\section{Optimizing lattice vortex function\label{optimize}}
In this section, we develop a theory to optimize the lattice analogy of the vortex function \cite{ledwith2023vortexability} for a given Chern-insulator tight-binding model.
We define an indicator that measures the distance from the ideal Chern insulator and discuss its optimization.
In addition to the unimodular coordinate transformation, we consider virtual lattice reconstruction of the internal sublattice position to optimize the indicator.
This degree of freedom serves as a discrete analog of the non-uniformity in the spatial metric and magnetic field in a continuous model.

\subsection{Notation}

Throughout the paper, we consider a tight-binding model given by:
\begin{align}
    \hat{H}=\sum_{\bm{R},\bm{R}'}\sum_{a,a'}c^\dagger_{\bm{R},a}H_{(\bm{R},a),(\bm{R}',a')} c_{\bm{R}',a'}, \label{tight}
\end{align}
where $\bm{R}=(X,Y)$ and $a,a'=1,2,\cdots,n_a$ denote the unit cell vector and intracell atomic orbital, respectively.
Here, $c$ represents the electron creation operator at the position $(\bm{R},a)$, and $H$ is the matrix representation of the quadratic Hamiltonian.
We consider a discrete translation-invariant two-dimensional system with an infinite number of unit cells without boundaries, or with a finite number of unit cells under periodic boundary conditions.
The former boundary condition is employed for formulations, whereas the latter is mainly utilized for numerical implementations.
It is useful to introduce the Fourier transform to rewrite the translation-invariant Hamiltonian.
Typically, there are two types of Fourier transforms for the finite case:
\begin{align}
    &c^{\dagger}_{\bm{k},a}=\frac{1}{\sqrt{V}}\sum_{\bm{R}}e^{i\bm{k}\cdot\bm{R}}c^{\dagger}_{\bm{R},a},\label{indep}\\
    &C^{\dagger}_{\bm{k},a}=\frac{1}{\sqrt{V}}\sum_{\bm{R}}e^{i\bm{k}\cdot\bm{R}+\bm{r}_a}c^{\dagger}_{\bm{R},a},\label{position-dependent}
\end{align}
where $\bm{k}$ is the crystal momentum, $V$ is the number of unit cells, and $\bm{r}_a=(x_a,y_a)$ is the intracell sublattice position of the atomic orbital $a$. 
Note that the values of momentum-dependent quantum geometric tensor (Berry curvature and quantum metric tensor) can depend on the notation, while the Chern number does not.
Given the crucial role of sublattice positions, which are treated separately from the unit-cell vector, we consistently employ the notation (\ref{indep}) throughout the paper. The Bloch Hamiltonian matrix $H_{\bm{k}}$ is defined by the following equation:
\begin{align}
    \hat{H}=\sum_{\bm{k}}\sum_{a,b}c^\dagger_{\bm{k},a}[H_{\bm{k}}]_{a,b}
    c_{\bm{k},b}.
\end{align}
For the infinite system without boundaries, the Bloch Hamiltonian is identical to the infinite volume limit of the finite system with periodic boundaries.
In the position-dependent notation (\ref{position-dependent}), the Bloch Hamiltonian is expressed as
\begin{align}
    H'_{\bm{k}}=D_{\bm{k}}H_{\bm{k}}D^{\dagger}_{\bm{k}},
\end{align}
where $D_{\bm{k}}=\mathrm{diag}(\cdots,e^{-i\bm{k}\cdot\bm{r}_a},\cdots)$.
While $H_{\bm{k}}$ is always periodic in the Brillouin zone, $H'_{\bm{k}}$ is periodic only upto a unitary transformation:
\begin{align}
    H'_{\bm{k}+\bm{G}}=D_{\bm{G}}H'_{\bm{k}}D^{\dagger}_{\bm{G}},
\end{align}
where $\bm{G}$ is a reciprocal lattice vector.

\subsection{Vortexable band with discretized vortex function}
Reference \cite{ledwith2023vortexability} introduced the concept of the vortexable band along with its associated vortex function.
Here, we propose a discrete version within the tight-binding Hamiltonian formalism, where functions and operators correspond to vectors and matrices, respectively.
Let $P$ be the projection matrix onto the vorexable band and $Z$ a diagonal matrix in the position basis $\{\ket{\bm{R},a}\}$, with its element $z(\bm{R},a)$ serving as a discrete counterpart to the vortex function.
For any state $\ket{\psi}$ in the vortexable band, i.e. $P\ket{\psi}=\ket{\psi}$, the vortex-function matrix $Z$ satisfies
\begin{align}
    Z\ket{\psi}=PZ\ket{\psi},
\end{align}
which is equivalently expressed as
\begin{align}
    ZP=PZP\notag\\
    \Leftrightarrow QZP=0,\label{vortexable}
\end{align}
where $Q=1-P$.
Our focus lies in determining the concrete form of $z(\bm{R},a)$. The trivial solution $Z=1$ satisfies Eq. (\ref{vortexable}) for any band structure but is not relevant to the physics of fractional Chern insulators. 
In the context of LLL physics, the vortex function is nothing but the complex coordinate $z=x+iy$ \cite{yoshioka-textbook-02,fradkin2013field}.
As in the case of the complex coordinate function, the $\pm 2\pi$ vortex is attached to the point $(\bm{R}_0,a_0)$ through the following procedure \cite{ledwith2023vortexability}:
\begin{align}
    [z(\bm{R},a)-z(\bm{R}_0,a_0)]\psi,
\end{align} 
where $\psi$ is the state in the vortexable band. At the many-body level, as described by Ref. \cite{ledwith2023vortexability}, the following relation holds:
\begin{align}
    f(\cdots, z(\bm{R}_i,a_i),\cdots)\ket{\Psi}=\mathcal{P}f(\cdots, z(\bm{R}_i,a_i),\cdots)\ket{\Psi},\label{many-body}
\end{align}
where $z(\bm{R}_i,a_i)$ is the vortex function for the particle $i$, $f$ is a holomorphic function of $z$'s, $\mathcal{P}$ is the many-body projection operator onto the vortexable band, and $\ket{\Psi}$ is a many-body state in the vortexable band, i.e. $\ket{\Psi}=\mathcal{P}\ket{\Psi}$.
The property (\ref{many-body}) enables us to extend the LLL physics to any vortexable band. In particular,  the Laughlin-like state with filling factor $\nu=1/(2s+1)$ is given by \cite{ledwith2023vortexability}
\begin{align}
    \ket{\Psi^{(2s)}}=\prod_{i<j}[z(\bm{R}_i,a_i)-z(\bm{R}_j,a_j)]^{2s}\ket{\Psi},
\end{align}
where $\ket{\Psi}$ is the fully filled state.

Note that a Chern band is not always a vortexable band.
For a given tight-binding model of a Chern insulator, our task is to optimize a function that resembles the vortex function according to certain criteria.
Broadening the scope of our definition, we also refer to such functions as vortex functions.
In the following, we formulate the optimization of the vortex function for a Chern band.

\subsection{Vortex function under translation invariance}
In this paper, we assume the translation invariance for the relative ``coordinate" defined by the vortex function:
\begin{align}
    z(\bm{R}_i,a_i)-z(\bm{R}_j,a_j)=z(\bm{R}_i+\bm{R},a_i)-z(\bm{R}_j+\bm{R},a_j),
\end{align}
where $\bm{R}$ is a lattice vector.
This equation suggests the following form for the vortex function:
\begin{align}
    Z&=\alpha~ (X+\sum_a\tilde{x}_aP_a)+\beta~ (Y+\sum_a\tilde{y}_aP_a)\notag\\
    &=\alpha~ X+\beta~ Y +\sum_a\gamma_aP_a,\label{vortexfunction}
\end{align}
where $\alpha,\beta,\gamma_a\in\mathbb{C}$ are parameters we aim to determine, and $P_a:=\ket{a}\bra{a}$ is the projection operator onto sublattice $a$.  Note that $\tilde{x}_a$ and $\tilde{y}_a$ may not correspond to the sublattice positions of the real system. Thus, the optimization process of the vortex function involves the virtual lattice reconstruction of the internal sublattice positions.
By definition, the following relation holds:
\begin{align}
    \begin{pmatrix}
        \mathrm{Re}\alpha&\mathrm{Re}\beta\\
        \mathrm{Im}\alpha&\mathrm{Im}\beta
    \end{pmatrix}
    \begin{pmatrix}
        \tilde{x}_a\\
        \tilde{y}_a
    \end{pmatrix}=
    \begin{pmatrix}
        \mathrm{Re}\gamma_a\\
        \mathrm{Im}\gamma_a
    \end{pmatrix}.\label{sublattice}
\end{align}

Remarkably, the virtual lattice rearrangement corresponds to the spatial non-linearity in continuous models \cite{ledwith2023vortexability, estienne2023ideal}.
Recently, the direct correspondence between the ideal Chern insulators and the Landau levels under the non-uniform metric and magnetic field has been studied.
This non-uniform nature is treated via the vortex function that is nonlinear in real-space coordinates \cite{ledwith2023vortexability, estienne2023ideal}.
The nonlinearity in the continuous models, in which infinite degrees exist in a unit cell, corresponds to the differences between the virtual and real sublattice positions. Hence, the nonlinearity is incorporated into the term $\sum_a \gamma_a P_a$ in Eq. (\ref{vortexfunction}).

\subsection{An indicator for vortexability}
Here, we formulate an indicator designed to quantify the vortexability.
From Eq. (\ref{vortexable}), 
our task is to minimize the operator $QZP$ in some sense.
A natural choice for an indicator is given by
\begin{align}
    &\mathrm{Tr}[(QZP)^{\dagger}QZP]/V=\mathrm{Tr}[PZ^*QQZP]/V\notag\\
    &=\mathrm{Tr}[PZ^*QZP]/V
    =\langle Z^*,Z\rangle\geq0,\label{ind1}
\end{align}
where $\langle\cdot,\cdot\rangle:= \mathrm{Tr}[P~\cdot~Q~\cdot~P]/V$.
This is an expression of the trace condition \cite{Roy-geometry-14,Jackson-Moller-Roy-15} for the vortex function (\ref{vortexfunction}).
However, minimizing Eq. (\ref{ind1}) may lead to the trivial solution discussed above. 
Equation (\ref{ind1}) can be divided into geometrical and topological parts:
\begin{align}
    \langle Z^*,Z\rangle&=\left(\langle \mathcal{R},\mathcal{R}\rangle+\langle \mathcal{I},\mathcal{I}\rangle\right)-i\left(\langle \mathcal{I},\mathcal{R}\rangle-\langle \mathcal{R},\mathcal{I}\rangle\right)\notag\\
    &=:G-T,
\end{align}
where $\mathcal{R}$ and $\mathcal{I}$ are the real and imaginary parts of the vortex function $Z$, respectively.
When we choose $ (\tilde{x}_a,\tilde{y}_a)=(x_a,y_a)$, under certain rotation and scale transformations, the first $(G)$ and second ($T$) terms become the momentum integral of the trace of the quantum metric tensor for the notation (\ref{position-dependent}) and the Chern number of the system, respectively \cite{Roy-geometry-14}.
For a general choice of the ($\tilde{x}_a,\tilde{y}_a$), they represent those of the system with lattice reconstruction. While lattice reconstruction alters the geometric term, it does not affect the topological term. 
A small value of $\langle Z^*,Z\rangle$ resulting from the tiny magnitudes of both the first and second terms corresponds to the trivial solution we aim to eliminate.
In ideal Chern insulators, the first and second terms are not small and cancel each other out. Instead of $\langle Z^*,Z\rangle$, we consider the following indicator:
\begin{align}
    I(Z):=\frac{\langle Z^*,Z\rangle}{\langle Z,Z^*\rangle}
    =\frac{G-T}{G+T}.
\end{align}
A small value of $I(Z)$ indicates the cancellation of the first and second terms of $\langle Z^*,Z\rangle$.
Due to the denominator, $I(Z)$ remains invariant under scale transformation of the parameter set, $\ket{\alpha}\rightarrow a\ket{\alpha}$, where $\ket{\alpha}=(\cdots,\alpha_i,\cdots)^T:=(\alpha,\beta,\gamma_1,\gamma_2,\cdots)^T$.
Note that the optimization of $I(Z)$ is equivalent to the optimization of the following indicator:
\begin{align}
    J(Z)=\frac{G-T}{T}=\frac{2I}{1-I},
\end{align}
under the assumption of a nontrivial Chern number, ensuring a nonzero denominator.
This equivalence arises from $dI/d\alpha_i=0$ implying $dJ/d\alpha_i=0$.
The choice of the indicator is a matter of taste, while $I(Z)$ is useful in the following formalism.

\subsection{Optimization of vortex function }
The optimal vortex function is obtained when the indicator $I(Z)$ is minimized, and this optimization process is described as an eigenvalue problem.
The numerator can be rewritten as
\begin{align}
    \langle Z^*,Z\rangle&=\bra{\alpha}
    \begin{pmatrix}
    \langle X,X\rangle&\langle X,Y\rangle&\cdots\langle X,P_b\rangle\cdots\\
    \langle Y,X\rangle&\langle Y,Y\rangle&\cdots\langle Y,P_b\rangle\cdots\\
    \vdots&\vdots&\vdots\\
    \langle P_a,X\rangle&\langle P_a,Y\rangle&\cdots\langle P_a,P_b\rangle\cdots\\
    \vdots&\vdots&\vdots
    \end{pmatrix}\ket{\alpha}\notag\\
    &=:\bra{\alpha} A\ket{\alpha}.\label{amat}
\end{align}
From Eqs. (\ref{ind1}) and (\ref{amat}), $A$ is an $(n_a+2)\times(n_a+2)$ positive semi-definite Hermitian matrix. 
Similarly, the denominator is given by
\begin{align}
    \langle Z^*,Z\rangle=\bra{\alpha^*} A\ket{\alpha^*}=\bra{\alpha} A^T\ket{\alpha}=\bra{\alpha} A^*\ket{\alpha}.
\end{align}
Note that $\ket{\alpha}=(0,0,1,1,1,\cdots)^T$ corresponds to the trivial solution $Z=1$ and is the zero mode of both $A$ and $A^*$.
In addition, if $A$ has a non-trivial zero mode $\ket{\alpha_0}$, $\ket{\alpha_0^*}$ is the zero mode of $A^*$. To avoid the indeterminate form, we add a regulator to $I(Z)$:
\begin{align}
    I_{\epsilon}(Z)=\frac{\bra{\alpha} (A+\epsilon1)\ket{\alpha}}{\bra{\alpha} (A^*+\epsilon1)\ket{\alpha}}=:\frac{\bra{\alpha} A_{\epsilon}\ket{\alpha}}{\bra{\alpha} A^{*}_{\epsilon}\ket{\alpha}},
\end{align}
where $\epsilon$ is a small but finite positive number.
Under this regularization, $I_\epsilon(Z)$ for the trivial solution $Z=1$ becomes unity, allowing us to focus on the nontrivial solution.
Thanks to the regulator, $A^*_\epsilon$ is a positive definite matrix. By applying the Cholesky decomposition $A^*_\epsilon=L_{\epsilon}L^{\dagger}_{\epsilon}$ with $L_{\epsilon}$ being a lower triangular matrix, the indicator can be rewritten as
\begin{align}
    I_{\epsilon}(Z)=\frac{\bra{\alpha} A_{\epsilon}\ket{\alpha}}{\bra{\alpha}L_{\epsilon}L^{\dagger}_{\epsilon}\ket{\alpha} }=\frac{\bra{\alpha}L_{\epsilon}[L^{-1}_{\epsilon} A_{\epsilon}(L^{\dagger}_{\epsilon})^{-1}]L^{\dagger}_{\epsilon}\ket{\alpha}}{\bra{\alpha}L_{\epsilon}L^{\dagger}_{\epsilon}\ket{\alpha} }.
\end{align}
Thus, the minimum of $I_{\epsilon}(Z)$ is given by the smallest eigenvalues of the following matrix
\begin{align}
    M_{\epsilon}:=L^{-1}_{\epsilon} A_{\epsilon}(L^{\dagger}_{\epsilon})^{-1}.
\end{align}
From the corresponding eigenvector $\ket{v}$, the optimal set of $\ket{\alpha}=(\alpha,\beta,\gamma_1,\gamma_2\cdots)$ is given by
\begin{align}
    \ket{\alpha}=(L^{\dagger}_{\epsilon})^{-1}\ket{v}.
\end{align}
Using the optimal $(\alpha,\beta)$, we define a new coordinate:
\begin{align}
    &\alpha X+\beta Y=X'+iY'\notag\\
    \Leftrightarrow&
    \begin{pmatrix}
        X'\\
        Y'
    \end{pmatrix}
    =
    t
    \begin{pmatrix}
        X\\
        Y
    \end{pmatrix}~\mathrm{with}~t:=
    \begin{pmatrix}
        \mathrm{Re}\alpha& \mathrm{Re} \beta\\
        \mathrm{Im}\alpha&\mathrm{Im}\beta
    \end{pmatrix}.\label{transform}
\end{align}
We normalize the parameters of the vortex function ($\ket{\alpha}$) by $|\det t|=1$. Under this normalization, the following holds in the new basis:
\begin{align}
    \langle Z^*,Z\rangle&=\langle X',X'\rangle+\langle Y',Y'\rangle+i(\langle X',Y'\rangle-\langle Y',X'\rangle)\notag\\
    &=\int\frac{d^2k}{(2\pi)^2}\mathrm{Tr}~g'(\bm{k})-\frac{C'}{2\pi},\label{fubini}
\end{align}
where $\bm{k}$ denotes the crytal momentum, $g'$ is the quantum metric tensor, and $C'$ is the Chern number.
We have introduced the momentum-space picture, which will be explained in the following subsection.
Under the optimization, $C'$ should be positive, whose absolute value coincides with the Chern number in the original frame.
The trace of the quantum metric tensor is not invariant under the coordinate transformation, and the integral is minimized in the above new coordinate.
The virtual sublattice position, $(\tilde{x}_a,\tilde{y}_a)$, is calculated using Eq.(\ref{sublattice}). 
Except for the optimization of the virtual sublattice position, the expression (\ref{fubini}) has often been discussed in various previous works \cite{Roy-geometry-14,Jackson-Moller-Roy-15,ledwith2023vortexability}.
These studies correspond to the optimization of the top-left block of Eq. (\ref{amat}).
In the context of superfluid density, the minimal quantum metric for optimal sublattice position was also discussed \cite{huhtinen2022revisiting,herzog2022many}.
This may be related to the optimization of the bottom-right block of Eq. (\ref{amat}).

We also note that the above optimization is related to the exploration of the $r$-ideal Chern band in continuous models \cite{ledwith2023vortexability, estienne2023ideal}.
In the Fourier notation (\ref{position-dependent}), the sublattice position is incorporated in the Fourier factor. The continuous model can be regarded as the lattice model with infinite sublattices. Thus, the optimization of sublattice positions in tight-binding models corresponds to the optimization of a function (mentioned as $r'$ and $F(r)$ in Refs. \cite{ledwith2023vortexability, estienne2023ideal}, respectively).

\begin{table*}[]
  \caption{Optimization of lattice vortex function in several models.}
  \label{table1}
  \centering
  \begin{tabular}{c|c|c|c|c|c}
    \hline
    Model & $n_{a}$& ($\alpha,\beta$) & ($\tilde{x}_a,\tilde{y}_a$)&$I(Z)$&$J(Z)$\\
    \hline \hline
    QWZ~\cite{Qi-Wu-Zhang-06} &2&($1,i$)&$(0,0)$,$(0,0)$&0.175&0.425\\
    checkerboard~\cite{Neupert-Santos-Chamon-Mudry-11}
    &2&($1,i$)& \textcolor{red}{($0.250,-0.250$),($-0.250$,$0.250$)}&0.0377&0.0783 \\
    square~\cite{Sun-Gu-Katsura-DasSarma-11} & 3 & $(1,$-$i)$ & $(0,0)$,$(0,0)$,$(0,0)$&0.230&0.596 \\
    ruby~\cite{Hu-Kargarian-Fiete-11,Wu-Bernevig-Regnault-12} & 6 & $(1.07$,-$0.54+0.31i)$  & 
    \textcolor{red}{\begin{tabular}{c}(-0.166,0.169),(0.168,0.336),(-0.669,-0.336),\\(0.167,-0.169),(-0.167,-0.337),(0.666,0.334)
    \end{tabular}}&0.0101&0.0203\\
    \hline
  \end{tabular}
\end{table*}
\subsection{Momentum-space representation of $I(Z)$}
In the following, we move onto the momentum-space picture.
An eigenstate of the one-particle Hamiltonian is described as a Bloch state:
\begin{align}
    \psi_{\bm{k},\mathcal{B}}(\bm{R},a)=\frac{1}{\sqrt{V}}e^{i\bm{k}\cdot \bm{R}}u_{\bm{k},\mathcal{B}}(a),
\end{align}
where $\bm{k}$ is the crystal momentum, $\mathcal{B}$ is the band index, and $u_{\bm{k},\mathcal{B}}(a)$ represents the periodic part of the Bloch state.
We use $\mathcal{O}$ for the occupied (Chern) bands and $\mathcal{U}$ for the unoccupied bands.
Then the projection operators onto these bands are given by
\begin{align}
    P&=\sum_{\bm{k},\mathcal{O}}\ket{\psi_{\bm{k},\mathcal{O}}}\bra{\psi_{\bm{k},\mathcal{O}}},\\
    Q&=1-P=\sum_{\bm{k},\mathcal{U}}\ket{\psi_{\bm{k},\mathcal{U}}}\bra{\psi_{\bm{k},\mathcal{U}}}
\end{align}
To express the matrix $A$ in momentum space, the following relation is useful:
\begin{align}
    \bra{\psi_{\bm{k},\mathcal{U}}}R_i\ket{\psi_{\bm{k},\mathcal{O}}}=\bra{u_{\bm{k},\mathcal{U}}}(\textcolor{red}{+}i\partial_{k_i})\ket{u_{\bm{k},\mathcal{O}}},\label{derivative}
\end{align}
where $R_{i=1,2}=X,Y$.
For numerical calculations, we also rely on the following expression:
\begin{align}
    \bra{\psi_{\bm{k},\mathcal{U}}}R_i\ket{\psi_{\bm{k},\mathcal{O}}}=\textcolor{red}{+}i \frac{
    \bra{\psi_{\bm{k},\mathcal{U}}}\partial_{k_i}H_{\bm{k}} \ket{\psi_{\bm{k},\mathcal{O}}}
    }{E_{\bm{k},\mathcal{O}}-E_{\bm{k},\mathcal{U}}},
\end{align}
where $H_{\bm{k}}$ is the Bloch Hamiltonian, and $E_{\bm{k},\mathcal{B}}$ is the energy dispersion of the band $\mathcal{B}$.
By using Eq. (\ref{derivative}), we obtain explicit formulas for each element of the matrix $A$:
\begin{align}
    \langle R_i,R_j\rangle&=\sum_{\mathcal{O},\mathcal{U}}\int\frac{d^2k}{(2\pi)^2}\bra{\partial_{k_i}u_{\bm{k},\mathcal{O}}}u_{\bm{k},\mathcal{U}}\rangle\langle u_{\bm{k},\mathcal{U}}
    \ket{\partial_{k_j}u_{\bm{k},\mathcal{O}}}\notag\\
    &=\int\frac{d^2k}{(2\pi)^2}\chi_{ij}(\bm{k}),\\
    \langle P_a,R_i\rangle&=(\langle R_i,P_a\rangle)^*\notag\\
    &=\sum_{\mathcal{O},\mathcal{U}}\int\frac{d^2k}{(2\pi)^2}
    u^*_{\bm{k},\mathcal{O}}(a)u_{\bm{k},\mathcal{U}}(a)\notag\\
    &~~~~~~~~~~~~~~~~~~~~\bra{u_{\bm{k},\mathcal{U}}}(\textcolor{red}{+}i\partial_{k_i})\ket{u_{\bm{k},\mathcal{O}}},\\
    \langle P_a,P_b\rangle&=\sum_{\mathcal{O},\mathcal{U}}\int\frac{d^2k}{(2\pi)^2}
    u^*_{\bm{k},\mathcal{O}}(a)u_{\bm{k},\mathcal{U}}(a)u^*_{\bm{k},\mathcal{U}}(b)u_{\bm{k},\mathcal{O}}(b)\notag\\
    &=\sum_{\mathcal{O}}\int\frac{d^2k}{(2\pi)^2}|u_{\bm{k},\mathcal{O}}(a)|^2\left[\delta_{ab}-|u_{\bm{k},\mathcal{O}}(b)|^2\right],
\end{align}
where $\chi_{ij}(\bm{k})$ is the quantum geometric tensor at momentum $\bm{k}$.

\begin{figure}[]
\begin{center}
 \includegraphics[width=8cm,angle=0,clip]{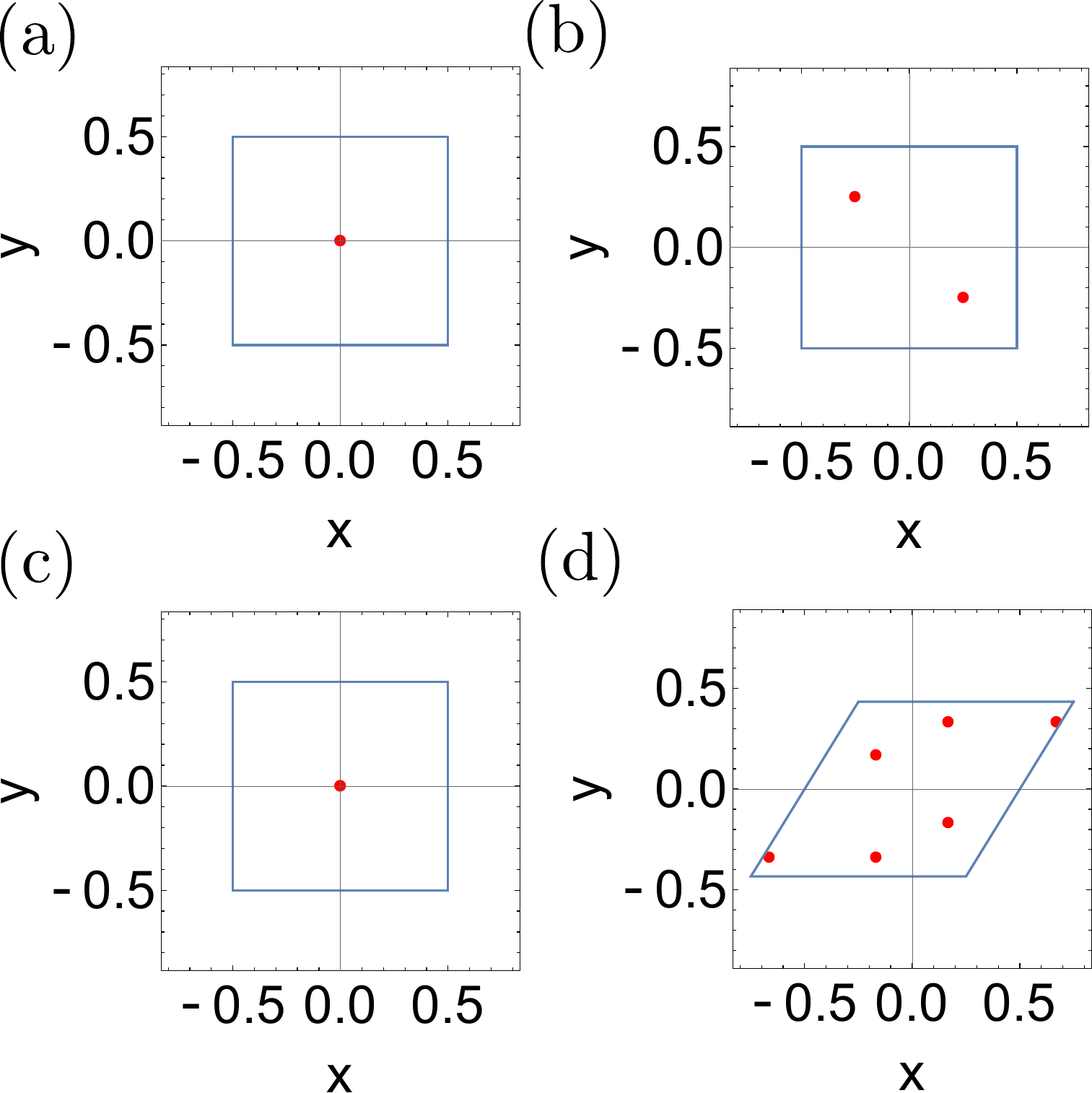}
 \caption{Optimal sublattice positions in (a)QWZ model~\cite{Qi-Wu-Zhang-06}, (b) checkerboard lattice model~\cite{Neupert-Santos-Chamon-Mudry-11}, (c) square lattice model~\cite{Sun-Gu-Katsura-DasSarma-11}, and (d) ruby lattice model~\cite{Hu-Kargarian-Fiete-11,Wu-Bernevig-Regnault-12}. Red points and blue lines represent the optimal sublattice points and unit cell boundaries, respectively.} 
 \label{fig1}
\end{center}
\end{figure}
\subsection{Examples}
We perform numerical calculations for several Chern insulators (Table \ref{table1}, Fig. \ref{fig1}). For each model, we focused on the lowest band with $|C|=1$.
Utilizing the momentum-space representation, we computed the matrix $A$, considering a system size of $V=32\times32$ and setting the regulator to $\epsilon=10^{-10}$. Details regarding the models and their parameters can be found in the Appendix.
As shown in Table \ref{table1}, the indicator $I(Z)$ strongly depends on the tight-binding model, even with the freedom to adjust sublattice positions. 
Notably, in the QWZ model and the square model, all sublattice positions converge at the origin, indicating a non-uniform magnetic field within the unit cell. 
This characteristic accounts for the relatively large values of  $I(Z)$ observed for these models.

\section{Construction of radially localized basis states from Dirac operator acting on momentum space\label{radialbasis}}
In this section, we present a theory concerning Landau-level-like radially localized basis states within a given Chern band with the Chern number $C$. We define these basis states as the eigenstates of an operator, which serves as a generalization of the angular momentum operator. Remarkably, this operator possesses at least $|C|$ exact zero modes, the origin of which is elucidated by the Atiyha-Singer index theorem in momentum space.
In connection with the zero mode, we introduce an inequality relevant to a topologically nontrivial band. Furthermore, we address a subtle distinction between our theory and prior work on the momentum-space Landau level \cite{Claassen-Lee-Thomale-Qi-Devereaux-15,Lee-Claassen-Thomale-17}.
Throughout the following discussion, we assume that  $Z$ has already been optimized, although the subsequent theory remains applicable even for a non-optimized $Z$.

\subsection{A construction of radially localized basis states}

In the lowest Landau level (LLL), the energy eigenstates are given by \cite{yoshioka-textbook-02,fradkin2013field}
\begin{align}
    \phi_m(z)=\frac{1}{\sqrt{2\pi 2^mm!}}z^m\exp(-|z^2|/4),\label{landau}
\end{align}
where $m$ is a non-negative integer representing the angular momentum, $z=x+iy$, and the magnetic length $l$ is set to unity.
The projection operator onto the LLL, denoted as $P_{\rm LLL}:=\sum_{m=0}^{\infty}\ket{\phi_m}\bra{\phi_m}$, gives the following relations:
\begin{align}
    &P_{\rm LLL}zP_{\rm LLL}z^{*}P_{\rm LLL}\ket{\phi_{m}}=2m\ket{\phi_{m}}.\label{LLLPPP}
\end{align}
This implies that the radially localized states in the LLL can be seen as eigenstates of the operator $P_{\rm LLL}zP_{\rm LLL}z^{*}P_{\rm LLL}$.
For $m=0$, Eq. (\ref{LLLPPP}) indicates that the LLL under a vortex attachment is orthogonal to $\phi_{m=0}$.
Note that this combination of operators can be considered as the angular momentum operator projected onto the LLL:
\begin{align}
    L_z=x(-i\partial_y)-y(-i\partial_x)=z\partial_z-z^*\partial_{z^*}.
\end{align}
For any states in the LLL, we have
\begin{align}
    z\partial_z[f(z)e^{-\frac{|z|^2}{4}}]&=zf'(z)e^{-\frac{|z|^2}{4}}-\frac{|z|^2}{4}e^{-\frac{|z|^2}{4}}\notag\\
    &=[\frac{1}{2}zP_{\rm LLL}z^*-\frac{1}{4}|z|^2][f(z)e^{-\frac{|z|^2}{4}}],\\
    z^*\partial_{z^*}[f(z)e^{-\frac{|z|^2}{4}}]&=-\frac{1}{4}|z|^2[f(z)e^{-\frac{|z|^2}{4}}],
\end{align}
where $f(z)$ is a polynomial of $z$.
This leads to:
\begin{align}
    P_{\rm LLL}L_zP_{\rm LLL}=\frac{1}{2}P_{\rm LLL}zP_{\rm LLL}z^{*}P_{\rm LLL}.\label{angular}
\end{align}

Now, we extend this observation to the Chern band.
We regard the eigenstates of $PZPZ^*P$ as Landau-level-like basis states:
\begin{align}
    &PZPZ^{*}P\ket{\phi_m}=\lambda_m\ket{\phi_m},\label{pzpzp}
\end{align}
where the eigenstates are in the Chern band ($P\ket{\phi_m}=\ket{\phi_m}$).
Due to the positive semi-definiteness of $PZPZ^*P$, we have $\lambda_m\geq0$.
We express $\phi_m$ using the Bloch states in the Chern band:
\begin{align}
    &\ket{\phi_m}=\sum_{\mathcal{O}}V\int\frac{d^2k}{(2\pi)^2}a^{\mathcal{O}}_m(\bm{k})\ket{\psi_{\bm{k},\mathcal{O}}},\label{expansion}\\
    &\sum_{\mathcal{O}}V\int\frac{d^2k}{(2\pi)^2}[a^{\mathcal{O}}_m(\bm{k})]^*a^{\mathcal{O}}_m(\bm{k})=1.
\end{align}
By substituting Eq.(\ref{expansion}) into Eq.(\ref{pzpzp}), we obtain
\begin{align}
    &\sum_{\mathcal{O},\mathcal{O}',\mathcal{O}''}\int\frac{d^2k}{(2\pi)^2}\ket{\psi_{\bm{k},\mathcal{O}}}\notag\\
    &\bra{u_{\bm{k},\mathcal{O}}}(\textcolor{red}{-}i\alpha \overleftarrow{\partial}_{k_x}\textcolor{red}{-}i\beta \overleftarrow{\partial}_{k_y}+\sum_a\gamma_aP_a)\ket{u_{\bm{k},\mathcal{O}'}}\notag\\
    &\bra{u_{\bm{k},\mathcal{O}'}}
    (\textcolor{red}{+}i\alpha^*\overrightarrow{\partial}_{k_x}\textcolor{red}{+}i\beta^* \overrightarrow{\partial}_{k_y}+\sum_a\gamma^*_aP_a)  \notag\\
    &(a^{\mathcal{O}''}_{m}(\bm{k})\ket{u_{\bm{k},\mathcal{O}''}}) \notag\\
    &=\lambda_m\sum_{\mathcal{O}}\int\frac{d^2k}{(2\pi)^2}a^{\mathcal{O}}_m(\bm{k})\ket{\psi_{\bm{k},\mathcal{O}}}\notag\\
    &\Leftrightarrow\notag\\
    &\hat{z}(\bm{k})\hat{z}^{\dagger}(\bm{k})\bm{a}_{m}(\bm{k})=\lambda_m \bm{a}_{m}(\bm{k}),\label{main}
\end{align}
where
\begin{align}
    \hat{z}(\bm{k})=&\textcolor{red}{-}\sum_{i}\alpha_i(-i\overrightarrow{\partial}_{k_i}+\hat{A}_{i}(\bm{k})),\notag\\
    \hat{z}^{\dagger}(\bm{k})=&\textcolor{red}{-}\sum_{i}\alpha^*_i(-i\overrightarrow{\partial}_{k_i}+\hat{A}_{i}(\bm{k})),\notag\\
    [\hat{A}_i(\bm{k})]_{\mathcal{O}\mathcal{O}'}=&-i\bra{u_{\bm{k},\mathcal{O}}}\partial_{k_i}\ket{u_{\bm{k},\mathcal{O}'}}\notag\\
    &\textcolor{red}{-}\sum_{a}\tilde{r}^i_au^*_{\bm{k},\mathcal{O}}(a)u_{\bm{k},\mathcal{O}'}(a).\label{berry}
\end{align}
The hat symbol denotes matrices in the occupied-band indices.
We use arrows over the derivatives to indicate their operational direction. The operators $\hat{z}(\bm{k})$ and $\hat{z}^{\dagger}(\bm{k})$ represent the vortex functions in momentum space.
The non-Abelian vector potential in momentum space, $\hat{A}_i(\bm{k})$, is identical to the Berry connection of the occupied bands (with the virtually optimized sublattice position) defined in the position-dependent convention of Fourier transform.

\subsection{Relation with momentum-space Landau level}
We here explain the major difference from the previous work about ``momentum-space Landau level (mLL)" \cite{Claassen-Lee-Thomale-Qi-Devereaux-15,Lee-Claassen-Thomale-17}.
In the context of the vortex function $Z$,
Refs. \cite{Claassen-Lee-Thomale-Qi-Devereaux-15,Lee-Claassen-Thomale-17} used $PX_{\mu}\eta^{\mu\nu}X_{\nu}P=PZZ^*P$, where $\eta^{\mu\nu}$ is a confinement metric, instead of $PZPZ^*P$ used in our paper.
The difference is given by
\begin{align}
    PZZ^*P-PZPZ^*P=PZQZ^*P~(\neq PZ^*QZP).
\end{align}
The topological and geometrical terms, both of substantial magnitude, nullify each other in $PZ^*QZP$, while they combine with the same sign in $PZQZ^*P$. In the momentum-resolved expression, 
a derivative equation within the mLL formalism is represented as follows \cite{Claassen-Lee-Thomale-Qi-Devereaux-15,Lee-Claassen-Thomale-17}:
\begin{align}
    &[z^{\dagger}(\bm{k})z(\bm{k})+(\eta^{\mu\nu}g_{\mu\nu}(\bm{k})-\Omega(\bm{k}))]a'_m(\bm{k})=\lambda'_m a'_m(\bm{k})\\
    \Leftrightarrow&
    [z(\bm{k})z^{\dagger}(\bm{k})+(\eta^{\mu\nu}g_{\mu\nu}(\bm{k})+\Omega(\bm{k}))]a'_m(\bm{k})=\lambda'_m a'_m(\bm{k}),
\end{align}
where $g$ and $\Omega$ are the quantum metric tensor and Berry curvature, respectively. It was assumed that only one band was occupied. We have used the commutation relation $[z^{\dagger}(\bm{k}),z(\bm{k})]=2\Omega(\bm{k})$.
Our equation (\ref{main}) differs from the one in the mLL formalism.
In general, they may give slightly different eigenstates.

In the following, we consider the condition under which these two formalisms coincide.
The following condition, 
\begin{align}
    \eta^{\mu\nu}g_{\mu\nu}(\bm{k})-\Omega(\bm{k})=0,\label{idealcond}
\end{align}
is often referred to as the trace condition or ideal condition for the fractional Chern insulator \cite{Roy-geometry-14,Jackson-Moller-Roy-15,ledwith2023vortexability}. This condition, stemming from $PZ^*QZP$, indicates vortexability. Thus, within the vortexable band:
\begin{align}
    &z^{\dagger}(\bm{k})z(\bm{k})a'_m(\bm{k})\notag\\
    &=
    [z(\bm{k})z^{\dagger}(\bm{k})+2\Omega(\bm{k}))]a'_m(\bm{k})=\lambda'_m a'_m(\bm{k}).\label{idealcase}
\end{align}
If we further assume that the Berry curvature is constant in momentum space, our result reproduces the mLL formalism.

\subsection{Physical interpretation of basis state: Dirac operator, zero modes, and index theorem}
The difference between the two formalisms is linked to the underlying physical interpretation.
The states in momentum-space Landau level are constructed to be eigenstates of a parabolic confinement potential $ZZ^*=x_{\mu}\eta^{\mu\nu}x_{\nu}$.
In our framework, the functions $\phi_m$ resemble the angular-momentum eigenstates in projected bands [Eq. (\ref{angular})].
Moreover, $\phi_0$ can be regarded as a  ``hole" state generated by the attachment of a vortex, as detailed in  Eq.(\ref{pzpzp}).
Interestingly, the eigenvalue for $\phi_0$, denoted as $\lambda_0$ in Eq. (\ref{main}), is rigorously zero for general Chern insulators. 
Here, we adopt a coordinate system where $(\alpha,\beta)=(1,i)$ and the Chern number $C$ is positive [see Eq. (\ref{transform})].
Let us consider the following Dirac operator in momentum space:
\begin{align}
    &\begin{pmatrix}
        0&\hat{z}(\bm{k})\\
        \hat{z}^{\dagger}(\bm{k})&0
    \end{pmatrix}
    =i\sum^2_{i=1} \gamma_i(\partial_{k_i}+i\hat{A}_i(\bm{k}))=i\slashed{D}(\bm{k}),\label{chirality}
\end{align}
where $\gamma_1=\textcolor{red}{+}\sigma_x,\gamma_2=\textcolor{red}{-}\sigma_y$ with $\sigma$'s being the Pauli matrices.
The left and right movers of the Dirac operator are defined as the negative and positive eigenstates of $\gamma_5:=-i\gamma_1\gamma_2=-\sigma_z$:
\begin{align}
    \gamma_5
    \begin{pmatrix}
        \psi_L\\
        0
    \end{pmatrix}=-
    \begin{pmatrix}
        \psi_L\\
        0
    \end{pmatrix},~
    \gamma_5
    \begin{pmatrix}
        0\\
        \psi_R
    \end{pmatrix}=+
    \begin{pmatrix}
        0\\
        \psi_R
    \end{pmatrix}.
\end{align}
Remarkably, the eigenequation of $i\slashed{D}(\bm{k})$ is related to Eq. (\ref{main}):
\begin{align}
    i\slashed{D}
    \begin{pmatrix}
        \bm{a}_m\\
        \pm\bm{b}_{m}
    \end{pmatrix}&=
    \begin{pmatrix}
        0&\hat{z}\\
        \hat{z}^{\dagger}&0
    \end{pmatrix}
    \begin{pmatrix}
        \bm{a}_m\\
        \pm\bm{b}_{m}
    \end{pmatrix}
    =\pm\sqrt{\lambda_m}
    \begin{pmatrix}
        \bm{a}_m\\
        \pm\bm{b}_{m}
    \end{pmatrix},
\end{align}
\begin{align}
    [i\slashed{D}]^2
    \begin{pmatrix}
        \bm{a}_m\\
        \bm{b}_{m}
    \end{pmatrix}
    =
    \begin{pmatrix}
        \hat{z}\hat{z}^{\dagger}&0\\
        0&\hat{z}^{\dagger}\hat{z}
    \end{pmatrix}
    \begin{pmatrix}
        \bm{a}_m\\
        \bm{b}_{m}
    \end{pmatrix}
    =
    \lambda_m
    \begin{pmatrix}
        \bm{a}_m\\
        \bm{b}_{m}
    \end{pmatrix},\label{doubled}
\end{align}
where $\bm{b}_{m}= z^{\dagger}\bm{a}_m/\sqrt{\lambda_m}$ for $\lambda_m\neq0$.
Essentially, the Landau-level-like basis emerges from the Dirac operator in momentum space.
This expression also sheds light on the existence of an exact zero mode.
According to the celebrated Atiyah-Singer index theorem, the following relation holds \cite{nakahara2018geometry}: 
\begin{align}
    \nu_R-\nu_L=-\int \frac{1}{n!}\mathrm{Tr}(\frac{\hat{F}}{2\pi})^n,
\end{align}
where $\nu_{R,L}$ is the number of zero modes with $\gamma_5=\pm 1$, 
$2n$ is the dimension of the manifold on which the Dirac operator lives, and $\hat{F}$ is the gauge field. The right-hand side corresponds to the $n$-th Chern number. In our case, the manifold corresponds precisely to the momentum space, mathematically a $2$-torus.
As previously mentioned, the Chern number $C$ is positive in our chosen coordinate system, thus leading to:
\begin{align}
    \nu_L-\nu_R=\int\frac{d^2k}{2\pi}\Omega(\bm{k})=C>0,
\end{align}
indicating that the momentum-space Dirac operator harbors at least $C$ left movers with $\lambda_m=0$. 
From Eq. (\ref{chirality}), the left mover corresponds to the first component of the spinor.
Consequently, the following holds for at least $C$ modes:
\begin{align}
       & i\slashed{D}
    \begin{pmatrix}
        \bm{a}_m\\
        0
    \end{pmatrix}=0\notag\\
    &\Leftrightarrow \hat{z}\hat{z}^{\dagger}\bm{a}_{m}=0,~\bm{b}_{m}=0,~\lambda_m=0.
\end{align}
For a natural configuration of the Berry curvature, one can assume $(\nu_R,\nu_L)=(0,C)$, thus resulting in $C$-fold degenerate zero modes.

Next, we consider the relationship between the different $\lambda_m$.
In general, $\hat{z}\bm{a}_m$ is not an eigenstate of $\hat{z}\hat{z}^{\dagger}$:
\begin{align}
    \hat{z}\hat{z}^{\dagger}(\hat{z}\bm{a}_m)=(\lambda_m+2\Omega(\bm{k}))(\hat{z}\bm{a}_m).
\end{align}
If the Berry curvature $\Omega(\bm{k})$ is momentum-independent, $\hat{z}\bm{a}_m$ becomes an eigenstate with the eigenvalue $\lambda_{m}+2\Omega$.
In such an ideal case, each level exhibits a $C$-fold degeneracy, prompting the introduction of two quantum numbers: the level index $n$ and the degree of degeneracy $l$, in place of $m$.
Then, the eigenspectrum is simply given by $\{\lambda_{n,l}=2\Omega~ n~|~n\in \mathbb{Z}_{\geq0}\}$.
These indices may work well for momentum-dependent $\Omega(\bm{k})$,  although the exact degeneracy remains preserved solely for zero modes.


Finally, we revisit the connection between our result and the mLL.
In the vortexable band, satisfying Eq. (\ref{idealcond}), we derive the following relations from Eq. (\ref{doubled}):
\begin{align}
    &a'_{n,l}=b_{n+1,l}=\frac{z^{\dagger}a_{n+1,l}}{\sqrt{\lambda_{n+1,l}}},\\
    &\lambda'_{n,l}=\lambda_{n+1,l}.\label{spect}
\end{align}
Equation (\ref{spect}) indicates that the difference in spectrum (except for zero modes) arises solely from deviations from the ideal condition (\ref{idealcond}).

\subsection{Inequality for a wavepacket from index theorem}
From the existence of zero mode(s) in $PZPZ^*P$, one can show a theorem about wavepackets in the Chern band by straightforward calculation.
For a two-dimensional infinite lattice system without boundary and with translation symmetry, the following theorem holds.
\\
\\
{\bf Theorem}~~Let $\ket{\phi}$ be a state in a Chern band with a positive (negative) Chern number $C$ and $P$ the projection operator onto the Chern band. Then, the following inequality holds:
\begin{align}
    \bra{\phi}r^2\ket{\phi}\geq
    \bra{\phi}xQx\ket{\phi}+\bra{\phi}yQy\ket{\phi}\pm 2\mathrm{Im}\bra{\phi}xQy\ket{\phi},\label{inequality}
\end{align}
where $(x,y)$ is the position operator, $r^2=x^2+y^2$, $Q=1-P$, and $-$ in $\pm$ corresponds to the negative $C$ case. For any given $P$, there exist(s) $\ket{\phi}$ such that the equality in the inequality (\ref{inequality}) is satisfied.
\\
\\
At first sight, this inequality seems to be very trivial because of the positive semi-definiteness of $PzPz^*P$ with $z=x+iy$. 
The crucial point is, however, the existence of a state that satisfies the equality condition for a fixed $P$. This theorem stems from the nontrivial band topology, shedding light on profound aspects of wavepacket behavior within Chern bands.

\subsection{Real-space calculation}
In the previous subsections, the construction of the basis set is reduced to a momentum-space calculation.
However, the notion of the Chern insulator can also be defined in disordered and quasi-crystalline systems, which can not be represented in the momentum-space picture.
Furthermore, even within periodic systems, the direct construction of the basis set may be useful.
For these reasons, we here consider the real-space calculation in finite-size systems.

The true vortex function should be defined in the infinite system without boundary. However, since the periodic boundary condition is incompatible with the vortex function $Z$, the expression $PZPZ^*P$ discussed above is not directly applicable when assuming periodic boundary conditions.
By construction, the radially localized basis set is localized near the origin, and the eigenstates of $PZPZ^*P$ in the infinite system are well approximated by the eigenstates of 
\begin{align}
    P_{\mathrm{reg}}P_{\rm PBC}ZP_{\rm PBC}Z^*P_{\rm PBC}P_{\mathrm{reg}},\label{region}
\end{align}
where $P_{\mathrm{PBC}}$ is the projection operator onto the Chern band under the periodic boundary condition, and $P_{\mathrm{reg}}$ is the projection operator onto a finite region around the origin, significantly smaller than the entire system.
It is noteworthy that the zero modes $\lambda_m=0$ in $P_{\rm PBC}$ degenerate with a lot of zero modes in $1-P_{\rm PBC}$ and $1-P_{\rm reg}$.  
Thus, we apply $P_{\rm reg}P_{\rm PBC}$ to zero modes of Eq. (\ref{region}) in order to extract $\ket{\phi_m}$.

This approximation is also useful to calculate $\langle X_i, X_j\rangle$:
\begin{align}
    \langle X_i, X_j\rangle_{\rm reg}= \mathrm{Tr}[P_{\mathrm{reg}}P_{\rm PBC}X_iP_{\rm PBC}X_jP_{\rm PBC}P_{\mathrm{reg}}]/V_{\rm reg},
\end{align}
where $V_{\rm reg}$ is the number of unit cells in the projection region. For example, one can approximate the Chern number in real space:
\begin{align}
    C\simeq -2\pi i\left(\langle X, Y\rangle_{\rm reg}-\langle Y, X\rangle_{\rm reg}\right).
\end{align}
\\
\textcolor{red}{Note added: just for the calculation of states near the origin, $P_{\rm reg}$ in Eq. (\ref{region}) is not needed. However, without $P_{\rm reg}$, the other quasi-zero mode appears at the boundary due to the incompatibility between position operator and periodic boundary condition.}

\begin{figure}[]
\begin{center}
 \includegraphics[width=8.5cm,angle=0,clip]{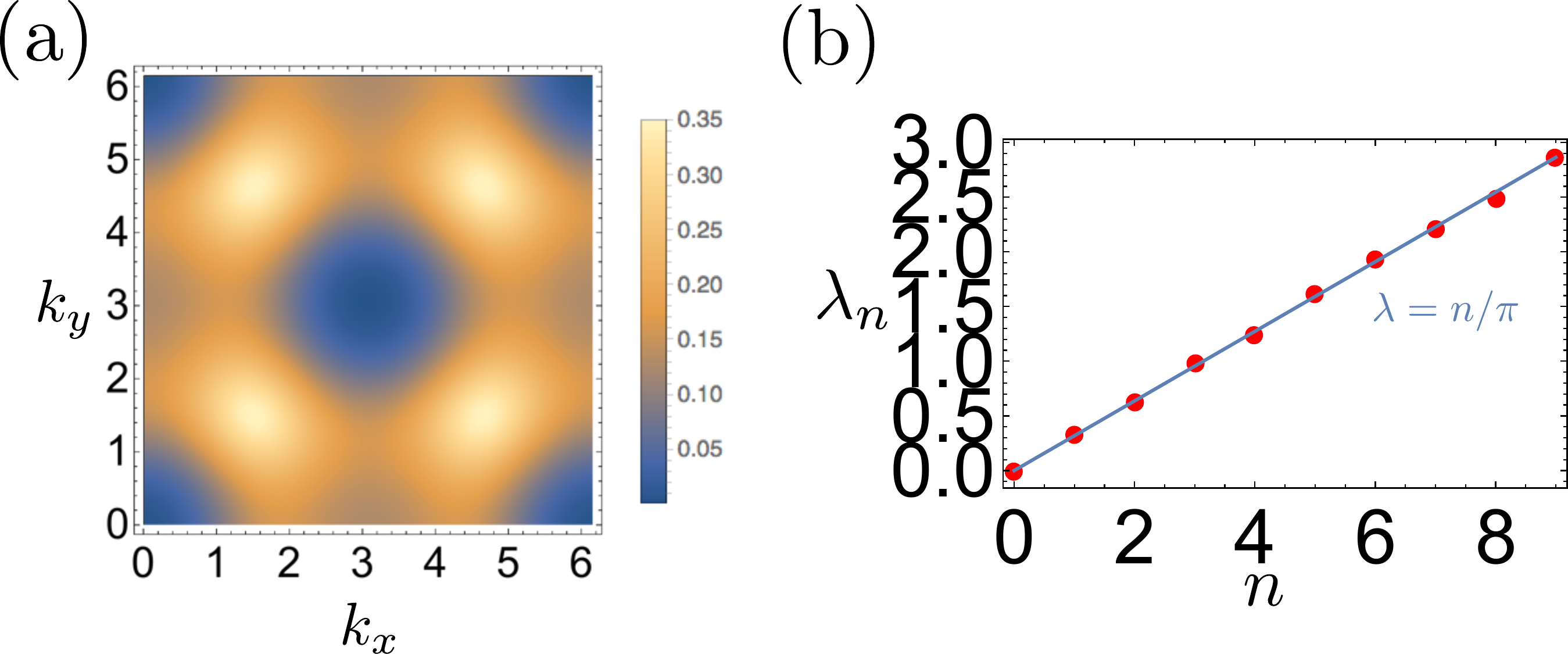}
 \caption{Landau-level-like basis states in checkerboard-lattice model~\cite{Neupert-Santos-Chamon-Mudry-11}. (a) Berry curvature distribution for optimized $\tilde{\bm{r}}_a$. (b) The ten smallest eigenvalues of $PZPZ^*P$. } 
 \label{fig2}
\end{center}
\end{figure}
\subsection{Example}
We here construct the basis states of the lowest Chern band $(C=1)$ in a checkerboard-lattice model \cite{Neupert-Santos-Chamon-Mudry-11}. We have used the vortex function determined in the previous section. 
For $(\alpha,\beta)=(1,i)$, the eigenvalue equation of the operator $z(\bm{k})z^{\dagger}(\bm{k})$ takes the form:
\begin{align}
    &\left[(-i\partial_{k_x}+A_x)^2+(-i\partial_{k_y}+A_y)^2-\Omega(\bm{k})\right]a_n(\bm{k})\notag\\&=\lambda_n a_n(\bm{k}).\label{eigen}
\end{align}
Here $\bm{A}$ and $\Omega$ are defined in sublattice-dependent notation (\ref{position-dependent}) for $\tilde{\bm{r}}_a$, as mentioned before.
Since $C=1$, we omit the quantum number $l$. 
 In the numerical calculation, we use the Hofstadter-type discrete model \cite{Hofstadter-76} with a momentum-dependent magnetic field to approximate the kinetic part in Eq. (\ref{eigen}).
The Brillouin zone is discretized into a $48\times48$ grid.
To evaluate the Berry curvature numerically, we use the Fukui-Hatsugai-Suzuki formula \cite{Fukui-Hatsugai-Suzuki-05} [Fig.\ref{fig2}(a)].

The smallest 10 eigenvalues of $z(\bm{k})z^{\dagger}(\bm{k})$ are depicted in Fig. \ref{fig2}(b).
These eigenvalues exhibit slight fluctuations around $\bar{\lambda}_n=2\bar{\Omega}n$, where $\bar{\Omega}=C/2\pi$ is the average of the Berry curvature on the Brillouin zone. 

\begin{figure}[]
\begin{center}
 \includegraphics[width=8.5cm,angle=0,clip]{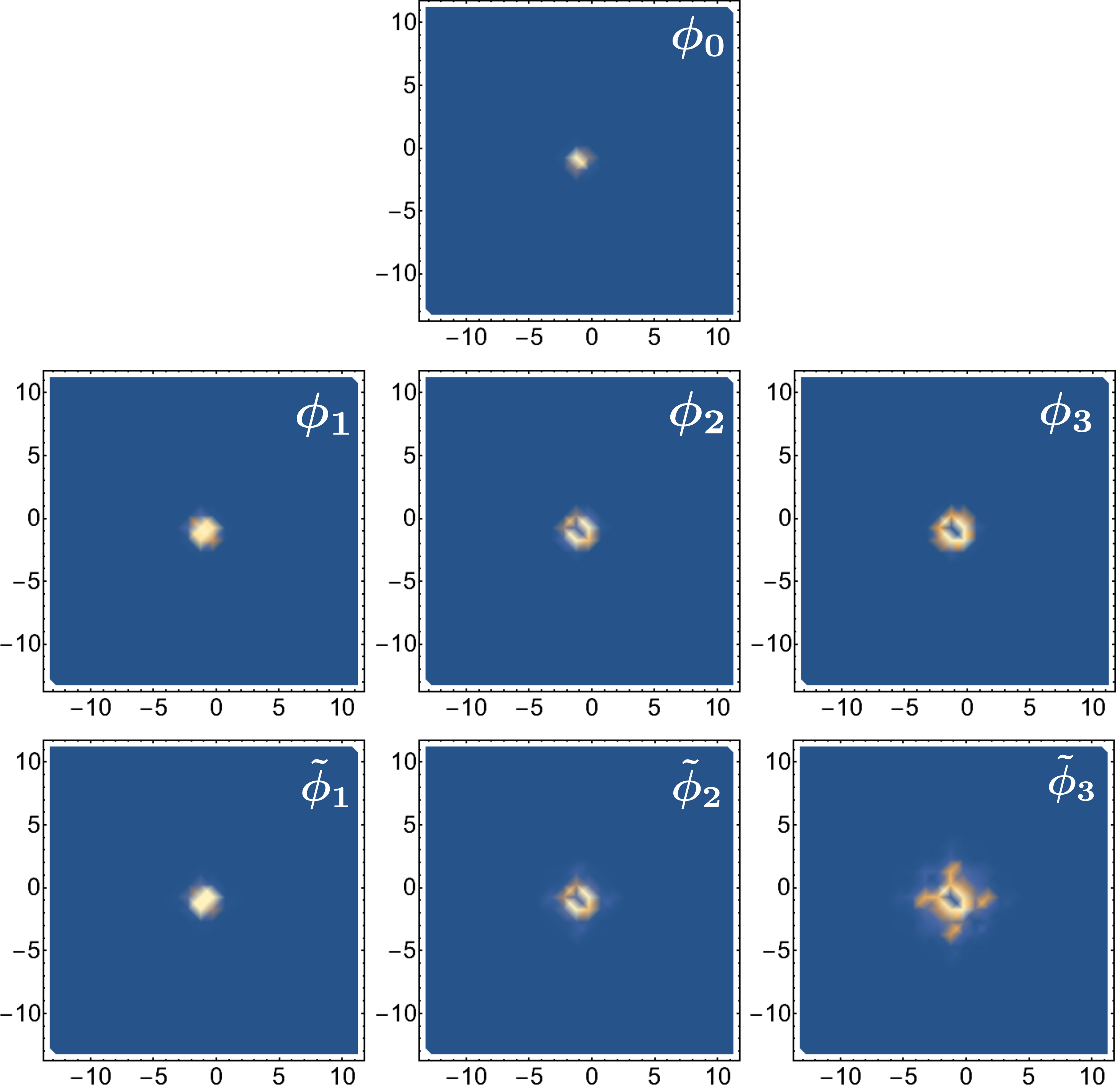}
 \caption{Real-space weight functions of $\ket{\phi_n}$ and $\ket{\tilde{\phi}_n}$.} 
 \label{fig3}
\end{center}
\end{figure}

For the real-space description of Landau-level-like basis states,  Eq. (\ref{region}) proves convenient.
In Fig. \ref{fig3}, we present the weight functions of $\ket{\phi_n}$ and $\ket{\tilde{\phi}_n}\propto (PZP)^n\ket{\phi_0}$.
Here, we have used $25\times25$ and $17\times17$ unit cells for $P_{\rm PBC}$ and $P_{\rm reg}$, respectively.
Evidently, these states exhibit minimal overlaps for large $n$.
Remarkably, a small $I(Z)$ does not guarantee the analytic properties of the basis states found in Eq. (\ref{landau}).

\section{Coherent-like states on von Neumann lattice\label{clsvnl}}
In this section, we extend the concept of coherent states from Landau levels to Chern insulators.
We further develop a theoretical framework to compare the LLL and Chern insulators in the context of coherent-like states.

In the LLL, the complex coordinate operators can be regarded as the creation and annihilation operators \cite{yoshioka-textbook-02,fradkin2013field}:
\begin{align}
    P_{\rm LLL}zP_{\rm LLL}=\sqrt{2} a^{\dagger},~
     P_{\rm LLL}z^*P_{\rm LLL}=\sqrt{2} a.
\end{align}
The $m=0$ state $\ket{\phi_{m=0}}$ in Eq. (\ref{LLLPPP}) is the vacuum of $a$.
Generally, coherent states for a given annihilation operator $a$ are defined as:
\begin{align}
    a\ket{\bar{\alpha}}=\bar{\alpha}\ket{\bar{\alpha}},
\end{align}
where $\bar{\alpha}$ is an arbitrary complex number, with the vacuum state corresponding to $\bar{\alpha}=0$.
It is well known that the coherent states form an overcomplete set. As suggested by von Neumann \cite{von2018mathematical} and proved later \cite{perelomov2002completeness, bargmann1971completeness}, however, a subset characterized by complex numbers on a periodic lattice, called von Neumann lattice, forms just a complete set. In the LLL context, these states are:
\begin{align}
    P_{\rm LLL}z^*P_{\rm LLL}\ket{\alpha_{mn}}&=(m\omega_x+n\omega_y)\ket{\alpha_{mn}},\label{vnl}
\end{align}
where $m,n\in\mathbb{Z}$, and $\omega_x,\omega_y\in\mathbb{C}$ satisfy Im$[\omega_x\omega_y]=1$.
Since $\{\ket{\alpha_{mn}}\}$ is not orthogonal, the biorthogonal bra vectors $\{\langle\!\langle \alpha_{mn}|\}$ should be properly defined. See Ref.\cite{ishikawa1992field} for details.
Due to translation invariance, the Fourier transform of Eq. (\ref{vnl}), denoted as a non-normalized vector $\ket{\alpha_{\bm{p}}}$, is particularly useful.
According to Ref.\cite{ishikawa1999field}, the following relation holds:
\begin{align}
    \bra{\alpha_{\bm{p}}}\alpha_{\bm{p}'}\rangle&=\alpha(\bm{p})\sum_{\bm{N}}(2\pi)^2\delta(\bm{p}-\bm{p}'-2\pi \bm{N}),\label{orthogonal}\\
    \alpha(\bm{p})&=[\beta(\bm{p})]^*\beta(\bm{p}),\\
    \beta(\bm{p})&=(2\mathrm{Im}\tau)^{\frac{1}{4}}e^{i\frac{\tau}{4\pi}p_y^2}\vartheta_1(\frac{p_x+\tau p_y}{2\pi}|\tau),
\end{align}
where $\bm{N}$ is an integer-valued vector, $\tau=-\omega_x/\omega_y$, and $\vartheta_1$ is the elliptic theta function. For example, $\tau=i$ corresponds to the square lattice.
The states created by normalizing $\ket{\alpha_{\bm{p}}}$ \footnote{Since $\ket{\alpha_{\bm{p}=\bm{0}}}$ is a null state, special treatment is required in the case of zero momentum. See Refs.\cite{ishikawa1992field,ishikawa1999field} for details.} and their Fourier transforms are natural generalizations of the Bloch and Wannier functions to the von Neumann lattice \cite{ishikawa1999field}.
The advantage of this expression lies in its ability to represent the physics of LLL on the Brillouin zone.

\begin{figure}[]
\begin{center}
 \includegraphics[width=8.5cm,angle=0,clip]{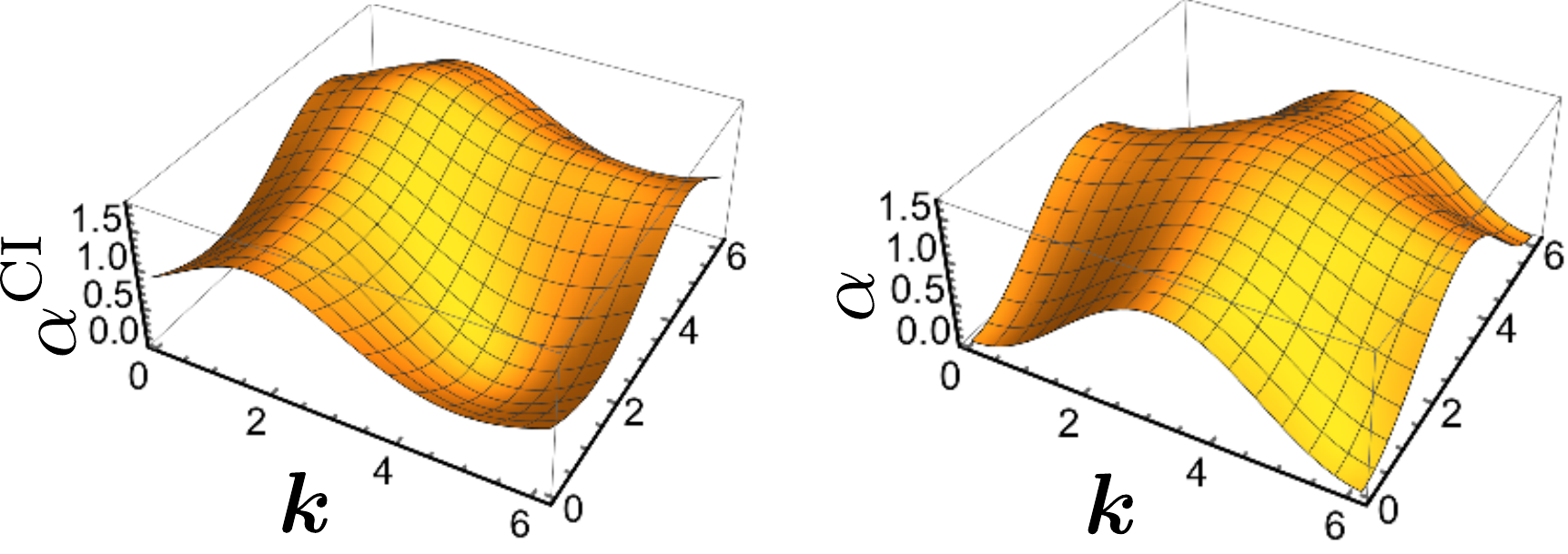}
 \caption{$\alpha(\bm{k})$ and $\alpha^{\rm CI}(\bm{k})$ for checkerboard-lattice model \cite{Neupert-Santos-Chamon-Mudry-11}.} 
 \label{fig4}
\end{center}
\end{figure}

Now, we generalize the notion of coherent states on the von Neumann lattice [Eq.(\ref{vnl})] to Chern insulators.
While $m=0$ state(s) in Eq. (\ref{expansion}) may not be eigenstates of an annihilation operator, it is still the exact zero mode of $PZ^*P$. Thus, they can be interpreted as a generalization of coherent states in the LLL.
In momentum-space, we have:
\begin{align}
    \hat{z}^{\dagger}(\bm{k})\bm{a}_{0}(\bm{k})=0.
\end{align}
For an arbitrary two-dimensional real vector $\bm{d}$, $\bm{a}_{\bm{d}}(\bm{k}):=e^{-i\bm{k}\cdot\bm{d}}\bm{a}_{0}(\bm{k})$ satisfies:
\begin{align}
    \hat{z}^{\dagger}(\bm{k})\bm{a}_{\bm{d}}(\bm{k})=\textcolor{red}{(\alpha^*d_x+\beta^*d_y)}\bm{a}_{\bm{d}}(\bm{k}).
\end{align}
Hence, the generalization of the coherent state is defined as:
\begin{align}
    &\ket{\alpha^{\rm CI}_{\bm{d}}}=\sum_{\mathcal{O}}V\int\frac{d^2k}{(2\pi)^2}e^{-i\bm{k}\cdot \bm{d}}a^{\mathcal{O}}_0(\bm{k})\ket{\psi_{\bm{k},\mathcal{O}}},\\
    &PZ^*P\ket{\alpha^{\rm CI}_{\bm{d}}}=\textcolor{red}{(\alpha^*d_x+\beta^*d_y)}\ket{\alpha^{\rm CI}_{\bm{d}}}.\label{chernvnl}
\end{align}
For $\bm{d}=\bm{R}$, 
Eq.(\ref{chernvnl}) mimics Eq. (\ref{vnl}), indicating that the unit cell vectors of the tight-binding model resemble a von Neumann lattice.
The Fourier transform of $\ket{\alpha^{\rm CI}_{\bm{R}}}$ is given by
\begin{align}
    \ket{\alpha^{\rm CI}_{\bm{k}}}=\sqrt{V}\sum_{\mathcal{O}}a^{\mathcal{O}}_{0}(\bm{k})\ket{\psi_{\bm{k},\mathcal{O}}}.
\end{align}
This vector satisfies the following orthogonality condition:
\begin{align}
    \bra{\alpha^{\rm CI}_{\bm{k}}} \alpha^{\rm CI}_{\bm{k}'}\rangle&=\alpha^{\rm CI}(\bm{k})~\delta_{\bm{k},\bm{k}'},\label{chernorthogonal}\\
    \alpha^{\rm CI}(\bm{k})&=V\sum_{\mathcal{O}}[a^{\mathcal{O}}_{0}(\bm{k})]^*a^{\mathcal{O}}_{0}(\bm{k}).
\end{align}
This equation serves as a generalization of Eq. (\ref{orthogonal}).
In Fig. \ref{fig4}, we compare $\alpha^{\rm CI}(\bm{k})$ for the checkerboard-lattice model \cite{Neupert-Santos-Chamon-Mudry-11} with $\alpha(\bm{k})$ in Eq. (\ref{orthogonal}). Depending on the gauge choice, the zero point of $\alpha^{\rm CI}(\bm{k})$ does not have to coincide with that of $\alpha(\bm{k})$. Except for the difference due to translation, both are similar in shape.
Again, we emphasize that this formalism enables us to compare the LLL and the given Chern insulator directly in the Brillouin zone.

We remark on a relationship between the coherent-like states $\ket{\alpha}$ and the radially localized states $\ket{\phi_m}$.
The operator $PZ^*P$ is expressed as a semi-infinite non-Hermitian matrix in the radially localized basis set:
\begin{align}
    PZ^*P&=\sum_{m,n}\ket{\phi_m}A_{m,n}\bra{\phi_n},
\end{align}
where $A_{m,n}=\bra{\phi_m}Z^*\ket{\phi_n}$.
For instance, the matrix representation for the LLL is given by
\begin{align}
    A=\sqrt{2}
    \begin{pmatrix}
     0&1&0&0&\cdots\\
     0&0&\sqrt{2}&0&\cdots\\
     0&0&0&\sqrt{3}&\cdots\\
     \vdots&\vdots&\vdots&\ddots&\ddots
    \end{pmatrix}.
\end{align}
If we regard $A$ as the non-Hermitian Hamiltonian on the semi-infinite lattice $\{m\}$, the coherent-like states are the localized non-Hermitian modes.
These modes are closely related to the non-Hermitian skin effect \cite{okuma2023non}, except for the root factor.

We also remark that similar works based on coherent states have already been conducted in Refs. \cite{qi2011generic,lee2013pseudopotential,jian2013crystal,lee2014lattice,lee2015exact}.
Due to the difference in construction, the relationship between coherent states in previous studies and ``coherent-like" states in our paper is not trivial. While the previous works determine the maximally localized coherent state in some way, our coherent-like state is uniquely defined as the zero mode(s) of $PZ^*P$ if the vortex function $Z$ is once determined. 
Note that in the numerical calculation of the FCI, the Wannier functions that are localized in one direction and delocalized in the other direction are often used \cite{qi2011generic}.
In terms of coherent states and coherent-like states, they may be interpreted as infinitely squeezing states in one direction.

\section{Discussion and Summary}
Here, we discuss the challenges we will address in the future. 
The two types of basis sets discussed in the latter part of the paper can be expected to be useful for many-body calculations of fractional Chern insulators. 
For example, the construction of composite fermions using coherent-like states on von Neumann lattice is intuitive.
In the Chern band, the vortex function and its Hermitian conjugate are written in a biorthogonal manner:
\begin{align}
    PZP&=\sum_{\bm{R}}Z_{\bm{R}}|\alpha^{\rm CI}_{\bm{R}}\rangle\!\rangle\bra{\alpha^{\rm CI}_{\bm{R}}},\notag\\
    PZ^*P&=\sum_{\bm{R}}Z^*_{\bm{R}}|\alpha^{\rm CI}_{\bm{R}}\rangle \langle\!\langle\alpha^{\rm CI}_{\bm{R}}|,
\end{align}
where $Z_{\bm{R}}=\alpha R_x+\beta R_y$ is the vortex function defined only on the lattice vectors.
Thus, the vortex function projected onto the Chern band is diagonal on the coherent-like basis. 
In this sense, the flux attachment is expected to be performed in an intuitive manner.

We are also interested in the correspondence between optimal vortex function and the stability of the fractional Chern insulator. While the virtual sublattice position rearrangement optimizes the kinetic part, its effect on the interaction part is very nontrivial.
We plan to investigate these points using the basis sets defined in the latter part.

Another interesting direction is the applications of our results to fractional physics in general topological insulators \cite{stern2016fractional}.
Actually, the $\mathbb{Z}_2$ version of the index theorem was studied in Ref. \cite{fukui2009z2}.
Thus, the latter part of our paper can be generalized to $\mathbb{Z}_2$ topological insulators.
In two-dimensional $\mathbb{Z}_2$ topological insulators, the Chern number is trivial, which means that $\nu_L=\nu_R$. However, the Dirac operator in momentum space (\ref{chirality}) has an odd number of zero-energy Kramers doublet when the $\mathbb{Z}_2$ index is nontrivial.
In such cases, the parity of $\nu_L$ (also $\nu_R$) becomes odd. Owing to the time-reversal symmetry, both the operators $PZ^*P$ and $PZP$ have protected zero modes.
These operators obey transpose-type time-reversal symmetry in non-Hermitian physics (see Appendix).
The above discussion indicates that the concepts in our paper can be relevant in the field of fractional topological insulators.

In this paper, we develop a theory of constructing the lattice vortex function and two types of basis sets for a given Chern insulator.
In the first half, we first define an indicator $I(Z)$ to quantify the vortexability.
We formulate the optimization of $I(Z)$ in the context of an eigenvalue problem.
Based on this theory, we determine the optimal vortex functions of several Chern-insulator models.
In the second half, we provide a theory of constructing the radially localized basis set and coherent-like basis set.
As a byproduct, we find the physics whose origin is the Atiyah-Singer index theorem for a Dirac operator on momentum space.

\acknowledgements
I thank Tomonari Mizoguchi \textcolor{red}{and Avedis Neehus} for the fruitful discussions.
This work was supported by JSPS KAKENHI Grant No.~JP20K14373 and No.~JP23K03243.
\\
\appendix
\section{Models}
In this Appendix, we write down the Bloch Hamiltonians and their parameters used in the main text.
\subsection{Qi-Wu-Zhang (QWZ) model}
This model was proposed in Ref.~\cite{Qi-Wu-Zhang-06} and is also referred to as the Wilson-Dirac model. 
The Bloch Hamiltonian reads
\begin{align}
    &H(\bm{k})=\sin k_1 \sigma_x+\sin k_2 \sigma_y+(m-\cos k_1-\cos k_2)\sigma_z, \label{eq:QWZ_Bloch}
\end{align}
where $\sigma_{x,y,z}$ are the Pauli matrices. 
The reciprocal lattice vectors are given as
\begin{align}
     \bm{G}_1=2\pi(1,0),\bm{G}_2=2\pi(0,1).
\end{align}
In numerical calculations, we set $m=1$.

\subsection{Checkerboard model}
This model was studied in terms of the fermionic FCI \cite{Neupert-Santos-Chamon-Mudry-11}. 
The Bloch Hamiltonian and the reciprocal lattice vectors are given as
\begin{align}
    &H(\bm{k})=
    \begin{pmatrix}
    2t_2(\cos k_1-\cos k_2)&t_1f^*(\bm{k})\\
    t_1f(\bm{k})&-2t_2(\cos k_1-\cos k_2)\\
    \end{pmatrix},\\
    &\bm{G}_1=2\pi(1,0),\bm{G}_2=2\pi(0,1),
\end{align}
where
\begin{align}
    &f(\bm{k})=e^{-i\pi/4}\left[1+e^{i(k_2-k_1)}\right]+e^{i\pi/4}\left[e^{-ik_1}+e^{ik_2}\right],\notag\\
    &t_1=1,t_2=\sqrt{2}/2.
\end{align}

\subsection{Square-lattice model}
This model was studied in terms of the fermionic FCI \cite{Sun-Gu-Katsura-DasSarma-11}.
This model is a three-orbital model on a square lattice. 
The Bloch Hamiltonian and reciprocal vectors are given as
\begin{widetext}
\begin{align}
    H(\bm{k}) = 
    \begin{pmatrix}
    -2t_{\rm dd} (\cos k_1 + \cos k_2) + \delta & 2i t_{\rm pd} \sin k_1 & 2it_{\rm pd} \sin k_2 \\
     & 2t_{\rm pp} \cos k_1 -2t_{\rm pp}^\prime \cos k_2 & i \Delta \\
      (\mathrm{h.c.})& & 2t_{\rm pp} \cos k_2 -2t_{\rm pp}^\prime \cos k_1 \\
    \end{pmatrix},\bm{G}_1 = (2\pi,0),\bm{G}_2 = (0,2\pi).
\end{align}
\end{widetext} 
We adopt the parameters, 
$t_{\rm dd} = t_{\rm pd} = t_{\rm pp} = 1$, 
$\Delta = 2.8$, 
$\delta = -4 t_{\rm dd} + 2 t_{\rm pp} + \Delta 
- 2t_{\rm pp} \Delta /  (4t_{\rm pp} + \Delta)$,
$t_{\rm pp}^\prime = t_{\rm pp} \Delta/ (4t_{\rm pp} + \Delta)$.

\subsection{Ruby-lattice model}
This model was studied in terms of the fermionic FCI \cite{Hu-Kargarian-Fiete-11,Wu-Bernevig-Regnault-12}.
The Bloch Hamiltonian and reciprocal vectors are given as

\begin{widetext}
\begin{align}
&H(\bm{k})
= - \begin{pmatrix}
0  & & & & & \\
  \tilde{t}_1^\ast & 0 & & & & \\
  \tilde{t} & \tilde{t}_1^\ast e^{-i(K_1 + K_2)} & 0 & & & \\
  t_4 ( 1 + e^{i K_1}) & \tilde{t} & \tilde{t}_1^\ast e^{i K_1} & 0 & & \\
  \tilde{t}^\ast & t_4( 1 + e^{-i (K_1+K_2)}) & \tilde{t} & \tilde{t}_1^\ast & 0 & \\
  \tilde{t}_1e^{iK_1} & \tilde{t}^\ast & t_4 ( e^{iK_1} + e^{i (K_1+K_2)}) & \tilde{t} & \tilde{t}_1^\ast e^{i (K_1+K_2)} & 0 \\
  \end{pmatrix}
  + (\mathrm{h.c.}), \notag\\
  &\bm{G}_1 = (2\pi, 2\pi/\sqrt{3}),\bm{G}_2 = (0,4\pi /\sqrt{3}),
\end{align}
\end{widetext}
where
$\tilde{t} = t_r + it_i$, $\tilde{t}_1 = t_{1r} + i t_{1i}$, $K_1 = k_1$, $K_2 = -k_1/2 + \sqrt{3}k_2/2$.
We adopt the parameters $(t_r,t_i,t_{1r},t_{1i},t_4) = (1,1.2, -1.2, 2.6, -1.2)$.

\section{Numerical calculation of momentum-space Hofstadter-type model on torus}
In this section, we provide technical details of the numerical computation of the eigenvalue problem for the ``Hamiltonian" in Eq. (\ref{eigen}).
Since we are mainly interested in the low-energy spectrum, 
we approximate the energy spectrum of the Hamiltonian operator by that of the corresponding lattice model, which is a Hofstadter-type model.
Note that each lattice site represents a crystal momentum on the momentum-space torus and the gauge field $A$ represents the Berry phase. 
We choose the continuous gauge using the parallel transport in Ref.\cite{soluyanov2012smooth}.
Under this gauge, all discontinuities are imposed on the Brillouin-zone boundary. 
The Hofstadter-type discretization of Eq. (\ref{eigen}) is defined by the following matrix elements:
\begin{align}
    &H_{\bm{k},\bm{k}+dk_x (1,0)}=-\textcolor{red}{\frac{L_xL_y}{(2\pi)^2}}U_x(i,j),\notag\\
    &H_{\bm{k}+dk_x (1,0),\bm{k}}=-\textcolor{red}{\frac{L_xL_y}{(2\pi)^2}}U^{\dagger}_x(i,j),\notag\\
    &H_{\bm{k},\bm{k}+dk_y (0,1)}=-\textcolor{red}{\frac{L_xL_y}{(2\pi)^2}}U_y(i,j),\notag\\
    &H_{\bm{k}+dk_y (0,1),\bm{k}}=-\textcolor{red}{\frac{L_xL_y}{(2\pi)^2}}U^{\dagger}_y(i,j),\notag\\
    &H_{\bm{k},\bm{k}}=\textcolor{red}{\frac{L_xL_y}{(2\pi)^2}}\left[4-\omega(i,j)\right],
\end{align}
where $\bm{k}=(\frac{2\pi i}{L_x},\frac{2\pi j}{L_y})$, and $d\bm{k}=(\frac{2\pi}{L_x},\frac{2\pi}{L_y})$.
$U_{x,y}$ represent the Peiels phase in $x$ and $y$ directions, which are matrices when the multiple bands are occupied. $\omega_{i,j}:=\arg U_x(i,j)U_y(i+1,j)U^{\dagger}_x(i,j+1)U_y^{\dagger}(i,j)$ is proportional to the Berry curvature in sublattice-dependent notation (\ref{position-dependent}).
All we have to do is determine the Peiels phase from the Bloch wavefunctions.
In the following, 
we assume that the number of occupied bands is one and the gauge of $\ket{u_{\bm{k}}}$ ($\mathcal{O}$ is dropped, the notation (\ref{indep}) is adopted) is already determined. 
If we ignore the sublattice-position term in Eq. (\ref{berry}), the Peiels phases are given by the normalized overlaps:
\begin{align}
    &U_x(i,j)=\exp(i\arg\langle u_{\bm{k}}|u_{\bm{k}+dk_x (1,0)}\rangle),\notag\\
    &U_y(i,j)=\exp(i\arg\langle u_{\bm{k}}|u_{\bm{k}+dk_y (0,1)}\rangle).
\end{align}
In the calculation at the Brillouin-zone boundary, we use 
\begin{align}
    \ket{u_{2\pi,k_y}}=\ket{u_{0,k_y}},~
    \ket{u_{k_x,2\pi}}=\ket{u_{k_x,0}}.
\end{align}
Thus, the wavefunctions at $k_{x,y}=2\pi-dk_{x,y}$ and $k_{x,y}=2\pi$ are not continuously connected in general, as mentioned above.
Under the presence of the sublattice effects, the Peiels phases are modified as
\begin{align}
    &U_x(i,j)=\exp(i\left[\arg\langle u_{\bm{k}}|u_{\bm{k}+dk_x (1,0)}\rangle\textcolor{red}{-}dk_x x^{(\mathrm{loc})}_{\bm{k}}\right]),\notag\\
    &U_y(i,j)=\exp(i\left[\arg\langle u_{\bm{k}}|u_{\bm{k}+dk_y (0,1)}\rangle\textcolor{red}{-}dk_yy^{(\mathrm{loc})}_{\bm{k}}\right]),
\end{align}
where
\begin{align}
    x^{(\mathrm{loc})}_{\bm{k}}=\bra{u_{\bm{k}}}(\sum_{a}\tilde{x}_aP_a)\ket{u_{\bm{k}}},\notag\\
    y^{(\mathrm{loc})}_{\bm{k}}=\bra{u_{\bm{k}}}(\sum_{a}\tilde{y}_aP_a)\ket{u_{\bm{k}}}.
\end{align}

\section{Transpose-type time-reversal symmetry}
In terms of non-Hermitian time-reversal symmetry, $PZ^*P$ (and $PZP$) has an interesting aspect. If the projection operator obeys a Hermitian time-reversal symmetry, $TP^*T^{-1}=P$, and $[T, Z^*]=0$, the following $\it{traspose}$-type time-reversal symmetry \cite{ksus} holds:
\begin{align}
    T(PZ^*P)^TT^{-1}=PZ^*P.
\end{align}
This can be checked by the following calculation:
\begin{align}
&T(PZ^*P)^{T}T^{-1}=TP^TZ^*P^TT^{-1}=TP^*Z^*P^*T^{-1}\notag\\
&=[TP^*T^{-1}][TZ^*T^{-1}][TP^*T^{-1}]=PZ^*P.
\end{align}
For $TT^*=-1$, this symmetry indicates the presence of non-Hermitian Kramers pair \cite{ksus}.
In our case, if $\ket{\alpha^{\rm CI}_{\bm{R}}}$ and $\langle\!\langle\alpha^{\rm CI}_{\bm{R}}|$ are right and left eigenstates of $PZ^*P$, $T|\alpha^{\rm CI}_{\bm{R}}\rangle\!\rangle^*$ becomes an right eigenstate of $PZ^*P$.
Unlike the Hermitian Kramers pair, the spatial distributions of these two states are completely different.
The latter assumption, $[T, Z^*]=0$, is natural in condensed matter in which $T$ represents a spin operator commuting with position operators.

\bibliography{FCI}
\end{document}